# Universal Quadratic Hierarchy Rule in Lepton Flavor Physics and Large Neutrino Mixing

E. M. Lipmanov
40 Wallingford Road # 272, Brighton MA 02135, USA

**Abstract**
Accurate values of solar and atmospheric neutrino oscillation large mixing parameters $\sin^2 2\theta_{12}$ and $\sin^2 2\theta_{23}$, and one-parameter neutrino mixing matrix with small deviations from Harrison-Perkins-Scott tribimaximal mixing, are obtained in terms of the new dimensionless constant $\alpha_o = e^{-5}$ in agreement with data. Basic lepton flavor deviation-from-extreme hierarchy-patterns for three different pairs of lepton mass-ratios and neutrino mixing parameters are shown to be approximate but distinct particular manifestations of one universal quadratic large-hierarchy rule (6) in lepton flavor physics.

## 1. Introduction

Since the recent suggestion of a new universal dimensionless constant

$$\alpha_o \equiv \exp(-5) \quad (1)$$

from hints of charged lepton (**CL)** data mass ratios[1] and their unification with quasi-degenerate (**QD**) neutrino mass ratios [1], it has been shown [2] that this constant may determine also the free in the electroweak theory interaction constants $\alpha$ and $\alpha_W$ at the pole values of the photon and W-boson propagators.

---

[1] The constant $\alpha_o$ may also describe the up- and down-quark mass ratios in quantitative analogy with the case of CL if the quark mass spectra are nearly geometrical [2].

In the present paper I consider the other lepton flavor problem of large neutrino mixing parameters in terms of deviation-from-maximal-mixing values by a straight analogy with the lepton deviation-from-degeneracy (DMD) hierarchy approach. A universal quadratic rule is quantitatively formulated for DMD- and deviation from maximal-neutrino mixing hierarchies in lepton flavor physics. It unveils physical analogy between the three deviation-from-extreme-hierarchies of dimensionless quantities in lepton flavor physics, and particularly leads to a connection between large neutrino mixing parameters from solar and atmospheric oscillations.

## 2. Quadratic rule for deviation-from-extreme hierarchies in lepton flavor physics

**A)** I suggest that the deviation-from-mass-degeneracy quantities of the charged lepton and neutrino mass$^2$ ratios, that are described in terms of the new parameter $\alpha_o$ in [1] and [2],

$$(m_\mu^2/m_e^2 - 1) \cong (1/2)(m_\tau^2/m_\mu^2 - 1)^2 \cong 2/\alpha_o^2, \qquad (2)$$

$$(m_2^2/m_1^2 - 1) \cong (1/2)((m_3^2/m_2^2 - 1)^2 \cong 2(5\alpha_o)^2, \qquad (3)$$

are two examples related to an independent of $\alpha_o$ DMD-hierarchy pattern in lepton flavor physics:

$$(m_\tau^2/m_\mu^2 - 1)^2 / (m_\mu^2/m_e^2 - 1) \cong 2, \qquad (4)$$

---

[2] $m_1 < m_2 < m_3$ denote the three neutrino masses.

$$(m_3^2/m_2^2 - 1)^2 / (m_2^2/m_1^2 - 1) \cong 2. \qquad (5)$$

Eq.(3) for the neutrino DMD-quantities is for the normal neutrino mass ordering; in case of reversed ordering, the ratios $(m_3/m_2)$ and $(m_2/m_1)$ should be interchanged.

In contrast to (2)-(3), the relations (4)-(5) do not depend on any dynamical constants or parameters, are symmetric in structure and may represent two different important realizations of a new probably universal flavor hierarchy rule of deviations from symmetry.

That rule as a universal DMD-hierarchy relation in flavor physics is given by:

$$[DMD(2)]^2/[DMD(1)] \cong 2, \qquad (6)$$

where $DMD(n)$, $n=1,2$, denote deviations from unity of the relevant lepton dimensionless quantities. This statement is a physical hypothesis testable by experimental data.

As an obvious test, the rule (6) should answer the specific neutrino problem of two experimentally large solar and atmospheric mixing parameters $\sin^2 2\theta_{12}$ and $\sin^2 2\theta_{23}$. In this case, the rule reads

$$(\sin^2 2\theta_{12} - 1)^2 / |\sin^2 2\theta_{23} - 1| \cong 2, \qquad (7)$$

as a new application of Eq.(6).

**B)** Consider the solutions of the nonlinear Eq.(6) in *exponential* form. There are two different types of solutions for large and small DMD-quantities:

**1)** In case of large DMD-values the obvious solution is $[DMD(2) = [2\exp(\chi)-1]$, $DMD(1) = [2\exp(2\chi)-1]$ with $\chi \gg 1$; it is for CL mass ratios: $DMD(2) \cong (m_\tau/m_\mu)^2$, $DMD(1) \cong (m_\mu/m_e)^2$, and comparison with experimental data yields [1] $\chi \cong 5$,

$$(m_\tau/m_\mu)^2 \cong 2/\alpha_o, \quad (m_\mu/m_e)^2 \cong 2/\alpha_o^2. \qquad (8)$$

**2)** In case of small DMD-values the solution is

$$DMD(2) \cong 2\Delta, \quad DMD(1) \cong 2\Delta^2, \quad \Delta \ll 1. \qquad (9)$$

There are two different applications of this solution:

**2a)** Solution for QD-neutrino mass ratios, $DMD(2) \cong [(m_3^2/m_2^2) - 1]$, $DMD(1) \cong [(m_2^2/m_1^2) - 1]$. In that case, the parameter $\Delta$ has a distinct physical meaning being the *observable* in neutrino oscillation experiments solar-atmospheric hierarchy parameter, $\Delta \cong r = (m_2^2 - m_1^2)/(m_3^2 - m_2^2)$,

$$(m_3^2/m_2^2) \cong \exp(2r), \quad (m_2^2/m_1^2) \cong \exp(2r^2). \qquad (10)$$

From comparison [1] with data $r \cong 5\exp(-5) = 5\alpha_o \cong 1/30$.

**2b)** Solution for 'large' neutrino mixing parameters, $DMD(2) = (\sin^2 2\theta_{12} - 1)$, $DMD(1) = (\sin^2 2\theta_{23} - 1)$. Comparison with *solar* neutrino experimental oscillation data [3, 4, 5] prompts the value $\Delta \cong \sqrt{\alpha_o}$,

$$(\sin^2 2\theta_{23} - 1) \cong -(1/2)(\sin^2 2\theta_{12} - 1)^2 \cong -2\alpha_o,$$

$$\sin^2 2\theta_{12} \cong \exp[-2\sqrt{\alpha_o}], \quad \sin^2 2\theta_{23} \cong \exp(-2\alpha_o). \qquad (11)$$

All three solutions of Eq.(6) are determined by the new constant $\alpha_o$, and have clear physical meaning similar to the CL hierarchy-solution (2) in terms of deviation-from-unity DMD-quantities.

From (11), a connection between the solar and atmospheric neutrino mixing parameters, directly measurable in accurate neutrino oscillation experiments, follows

$$\sin^2 2\theta_{23} = \exp[-0.5(\log \sin^2 2\theta_{12})^2]. \qquad (12)$$

So, if the solar neutrino oscillation mixing parameter is not maximal, nonmaximal mixing follows from (12) for the atmospheric oscillation neutrino mixing also.

**C)** By the solutions (8) and (11), large neutrino mixing parameters are related to the large charged lepton mass ratios in a symmetric way

$$\sin^2 2\theta_{12} \cong \exp(-2\sqrt{2}\, m_\mu/m_\tau), \qquad (13)$$

5$$\sin^2 2\theta_{23} \cong \exp(-2\sqrt{2}\, m_e/m_\mu). \qquad (14)$$

Relations (13) and (14) indicate a distinct reason of why the neutrino mixing parameters are large, but not maximal – since the charged lepton mass ratios $(m_\mu/m_\tau)$ and $(m_e/m_\mu)$ are small, but not zero[3]; and the atmospheric neutrino oscillation angle is closer to maximal than the solar one because of the large empirical CL mass-ratio hierarchy $m_e/m_\mu \ll m_\mu/m_\tau$.

The approximate[4] universal large-hierarchy DMD-rule (6) is independent of $\alpha_o$ and any outer parameters. Probably, it is a primary relation in lepton flavor physics. If so, lepton flavor physics of the three known generations is ruled by quadratic hierarchy of deviations-from-symmetry and neutrino-CL DMD-duality.

### 3. Neutrino mixing matrix

**A)** The two large neutrino mixing parameters are connected by relation (6). The solar mixing parameter from experimental data analysis [3] is,

$$(\tan^2\theta_{12})_{\exp} = (0.445 \pm 0.045). \qquad (15)$$

The quantitative result for *solar* neutrino mixing parameter (11) is given by

$$\sin^2 2\theta_{12} = \exp(-2\sqrt{\alpha_o}) \cong 0.8486, \tan^2\theta_{12} \cong 0.44, \qquad (16)$$

in agreement with the estimation (15).

---

[3] That answer to the problem of maximal neutrino mixing is similar to the other one [1] of why the neutrino mass spectrum should be quasi-degenerate, but not exactly degenerate.

[4] "…in the description of nature, one has to tolerate approximations, and that even work with approximations can be interesting and can sometimes be beautiful" - P. A. M. Dirac, Scientific autobiography, in *History of 20th Century Physics*, NY (1977).

From the solar value (15) and relation (12), the *atmospheric* neutrino mixing parameter should be

$$\sin^2 2\theta_{23} \cong 0.9873 \pm 0.006 \qquad (17)$$

in very good agreement with the value from (11):

$$\sin^2 2\theta_{23} = \exp(-2\alpha_o) \cong 0.9866. \qquad (18)$$

On the other hand, the obtained value (18) is in agreement with direct experimental data value for the atmospheric mixing parameter [3],

$$(\sin 2\theta_{23})_{\exp} \cong 1.01 \pm 0.02. \qquad (19)$$

This agreement supports the solution pattern (9) and (11).

Current experimental data analyses probably yield nonmaximal best-fit value for the atmospheric neutrino oscillation mixing angle $(\tan^2 \theta_{23})_{b-f} = 0.89$ from [4] and $(\tan^2 \theta_{23})_{b-f} = 0.82$ from [5] in agreement with the inference from (18): $\tan^2 \theta_{23} \cong 0.79$. As a result, the suggested hierarchy hypothesis of deviations from maximal mixing for two large neutrino mixing parameters seems to be supported by the considered above tests, and so probably may gain confirmation by new accurate experimental data.

**B)** If restricting to CP-conservation, the general form of neutrino mixing matrix $\nu_{e,\mu,\tau} = U\nu_{1,2,3}$ is given by [6]

$$U = \begin{pmatrix} C_{12} C_{13} & S_{12} C_{13} & S_{13} \\ -C_{23} S_{12} - C_{12} S_{13} S_{23} & C_{12} C_{23} - S_{12} S_{13} S_{23} & C_{13} S_{23} \\ S_{23} S_{12} - C_{12} C_{23} S_{13} & -C_{12} S_{23} - C_{23} S_{12} S_{13} & C_{13} C_{23} \end{pmatrix} \qquad (20)$$

where $C_{12} = \cos\theta_{12}$, $S_{12} = \sin\theta_{12}$, $S_{13} = \sin\theta_{13}$, etc. Here $\theta_{13}$

denotes the neutrino mixing angle from the CHOOZ reactor neutrino oscillation experiment $\nu_e \to \nu_\mu$ [3]:

$$(Sin^2 2\theta_{13})_{exp} = 0 \pm 0.05, \tag{21}$$

consistent with zero mixing angle $\theta_{13}$. Consider some comments to the point. If $\alpha_o = 0$ and $m_e$ - finite, it follows from (2) $m_\mu = \infty$, $m_\tau = \infty$; in that case the neutrino masses are exactly degenerate and so there are no physical oscillation transitions. Nevertheless, from (11) a relation for the large neutrino mixing parameters survived $Sin^2 2\theta_{23} = Sin^2 2\theta_{12} = 1$, formally meaning maximal mixing[5]. So, at small not zero value of the parameter $\alpha_o$ one should expect small *deviations* from maximal mixing for these two oscillation neutrino mixing parameters. But with three elementary particle generations, there is not analogous special relation for the neutrino mixing parameter $S_{13}$ and, therefore, the third neutrino mixing parameter should be small, $S_{13} \sim \alpha_o - 5\alpha_o$, suppressing oscillations of the type $\nu_e \to (\nu_\mu, \nu_\tau)$ with atmospheric-like oscillation length. As a probe, consider an example choice for the parameter $S_{13}$ (comp. Eq.(11)): $S_{13} = \sqrt{[(1 - Sin^2 2\theta_{23})(1 - Sin^2 2\theta_{12})]} = Cos 2\theta_{23} Cos 2\theta_{12} = (C_{23}^2 - S_{23}^2)(C_{12}^2 - S_{12}^2) \cong 2(\alpha_o)^{3/4} \cong 0.047$; in that case, $S_{13} = 0$ even if only one, atmospheric or solar, mixing parameter has maximal value, as in the HPS-example (24). More important is the inference that if both large neutrino mixing parameters are nonmaximal, $C_{23} \neq S_{23}$, $C_{12} \neq S_{12}$ as in (23), the small element of the neutrino mixing matrix should be finite $S_{13} \neq 0$.

---

[5] It resembles quantum perturbation theory in case of two-level-degeneration from a total of three ones.

**C)** In case $S_{13} \neq 0$, the one-parameter ($\alpha_o$) neutrino mixing matrix is predicted:

$$\begin{pmatrix} C_{12} & S_{12} & S_{13} \\ -C_{23} S_{12} & C_{12}C_{23} & S_{23} \\ S_{23} S_{12} & -C_{12} S_{23} & C_{23} \end{pmatrix} \cong \begin{pmatrix} 0.833 & 0.553 & S_{13} \\ -0.413 & 0.622 & 0.665 \\ 0.368 & -0.554 & 0.747 \end{pmatrix} \quad (23)$$

since the elements $S_{12}$ and $S_{23}$ of this matrix are determined by $\alpha_o$ by solution (11). Probably, the matrix element $S_{13}$ in (23) is small, $S_{13} \sim \alpha_o$, but not zero.

Compare the mixing matrix (23) with the widely discussed tribimaximal Harrison-Perkins-Scott (HPS) [7] mixing matrix with maximal atmospheric neutrino mixing

$$\begin{pmatrix} \sqrt{2}/\sqrt{3} & 1/\sqrt{3} & 0 \\ -1/\sqrt{6} & 1/\sqrt{3} & 1/\sqrt{2} \\ 1/\sqrt{6} & -1/\sqrt{3} & 1/\sqrt{2} \end{pmatrix} \quad (24)$$

The deviation of the atmospheric neutrino oscillation parameter $S_{23}$ (23) from the maximal-mixing-value (24) is small, but not zero ~ 6%. The deviation of the solar neutrino oscillation parameter $S_{12}$ (23) from the HPS value (24) is ~5%, while its deviation from maximal mixing value is ~28%.

Possible experimental tests of deviations of the neutrino mixing matrix from maximal mixing pattern in future neutrino telescope observations are considered in the literature, e.g. [8] and references therein.

## 4. Conclusions

The large solar and atmospheric oscillation neutrino mixing parameters $\sin^2 2\theta_{12}$ and $\sin^2 2\theta_{23}$ and the neutrino mixing matrix in case $s_{13} = 0$ are determined in terms of one small new universal physical parameter $\alpha_o$ from an interesting physical analogy with hierarchies of CL and neutrino mass ratios. The obtained one-parameter neutrino mixing matrix (23) is close to the tribimaximal mixing matrix of Harrison-Perkins-Scott, but is rather in better agreement with data. Special attention is paid to the suggested quantitative quadratic equation (6) for lepton DMD-hierarchies. The pair of deviations from maximal mixing of the neutrino oscillation mixing parameters follows as a new solution of Eq.(6). It agrees with experimental oscillation data analysis [3 - 5], and is a test of the discussed lepton flavor DMD-hierarchy rule.

The simplest phenomenological explanation of the empirical smallness of the neutrino oscillation hierarchy parameter *r*, together with large hierarchy of the deviations from maximal mixing, follows from an *analogy* with the known large CL mass-ratio hierarchy in the case of QD-neutrinos[6] – it is an analogy between all three known lepton basic deviation-from-extreme hierarchies.

In summery, the large mixing parameters of neutrino mass eigenstates ($\nu_1$, $\nu_2$, $\nu_3$) in the neutrino flavor eigenstates ($\nu_e$, $\nu_\mu$, $\nu_\tau$) are determined, and interpreted as closely related to the neutrino and CL mass-ratio patterns. The most interesting new physics result: the three known seemingly different charged lepton and QD-neutrino pressing problems of deviation-from-extreme-hierarchies are in reality three particular manifestations of *one unifying quadratic hierarchy-rule (6), as a probable conformity to natural laws in new lepton flavor physics.*

---

[6] Only in the QD-case the two neutrino DMD-quantities and neutrino DMD-hierarchy are simultaneously small as a result of large CL mass ratios and mass-ratio hierarchy [2].